\documentclass[3p,number,sort&compress]{elsarticle}
\usepackage{amsmath, amsthm, amssymb}

\def\<{\langle}
\def\>{\rangle}

\DeclareMathOperator{\Tr}{Tr}

\begin{document}
\begin{frontmatter}

\title{Nambu-Poisson Gauge Theory}

\author[CUP]{Branislav Jur\v co}
\ead{jurco@karlin.mff.cuni.cz}

\author[JUB]{Peter Schupp}
\ead{p.schupp@jacobs-university.de}

\author[JUB,CTU]{Jan Vysok\'y}
\ead{vysokjan@fjfi.cvut.cz}

\address[CUP]{Charles University in Prague, Faculty of Mathematics and Physics, Mathematical Institute,
Prague 186 75, Czech Republic}
\address[JUB]{Jacobs University Bremen, 28759 Bremen, Germany}
\address[CTU]{Czech Technical University in Prague, Faculty of Nuclear Sciences
and Physical Engineering, Prague 115 19, Czech Republic}

\begin{abstract}
We generalize noncommutative gauge theory using Nambu-Poisson structures to obtain a new type of gauge theory with higher brackets and gauge fields. The approach is based on covariant coordinates and higher versions of the Seiberg-Witten map. We construct a covariant Nambu-Poisson gauge theory action, give its first order expansion in the Nambu-Poisson tensor and relate it to a Nambu-Poisson matrix model.
\end{abstract}


\end{frontmatter}

\section{Introduction}\label{intro}

In this letter, we introduce a higher analogue of noncommutative (abelian) pure gauge theory. What we consider here is a deformation, in the presence of a background $(p+1)$-rank Nambu-Poisson tensor, of an abelian gauge theory with a $p$-form gauge potential, i.e., a $(p-1)$-gerbe connection. Our approach, for $p>1$, is similar to that of \cite{Madore:2000en} which deals with the more familiar case of $p=1$. 
A Nambu-Poisson gauge theory was pioneered by P.-M. Ho et al. in \cite{Ho:2008ve} as the effective theory of M5-brane for a large longitudinal $C$-field background in M-theory. Related work can be found in their papers \cite{Ho:2007vk,Ho:2008nn,Ho:2009zt}.

We formulate the theory independently of string/M-theory. Nevertheless, the motivation comes from M-theory branes; more explicitly from an effective DBI-type theory proposed for the description of multiple M2-branes ending on a M5-brane, where the Nambu-Poisson 3-tensor enters as a pseudoinverse of the 3-form field $C$ \cite{Jurco:2012yv,JSV1}. We develop the theory at a semiclassical level, briefly commenting on the issue of quantization at the end.

The paper is organized as follows: After discussing conventions in Sec.~\ref{Conv}, we introduce in Sec.~\ref{CovX} covariant coordinates, which transform nontrivially with respect to gauge transformations parametrized by a $(p-1)$-form, the gauge transformation being described in terms of a $(p+1)$-bracket arising from a background Nambu-Poisson $(p+1)$-tensor. Based on these covariant coordinates, we introduce Nambu-Poisson gauge fields in Sec.~\ref{NPGF}. In Sec.~\ref{via SW}, we construct Nambu-Poisson gauge fields explicitly, using a suitable generalization \cite{Chen:2010br,Jurco:2012yv,JSV1} of the Seiberg-Witten map~\cite{Seiberg:1999vs}, starting form an ordinary $(p-1)$-form gauge potential. We give explicit expressions for all components of the Nambu-Poisson field strength. In Sec.~\ref{action}, we give the corresponding (semiclassically) ``noncommutative'' action and its first order expansion in the Nambu-Poisson tensor.
Up to this order the the result is unambiguous, because quantum corrections from any type of quantization of the Nambu-Poisson structure will only affect higher orders. We conclude the letter by relating the action to (the semiclassical version of) a Nambu-Poisson matrix model.

We only briefly comment on deformation quantization of Nambu-Poisson structures in this letter. A satisfactory description of Nambu-Poisson noncommutative gauge theory beyond the semiclassical level will require a suitable analogue of Kontsevich's formality, solving in particular the deformation quantization problem for an arbitrary Nambu-Poisson structure.

\section{Conventions}\label{Conv}

We assume that $n$-dimensional space-time~$M$ is equipped with a rank $p+1$ Nambu-Poisson structure~$\Pi$, with $1 < p < n$.\footnote{The discussion could be extended to include also the well known case $p=1$, but for clarity and brevity we concentrate here on $p>1$ and refer to~\cite{JSV1} for $p=1$.} The corresponding Nambu-Poisson bracket is denoted by $\{\cdot,\ldots,\cdot\}$. In order to keep notation close to the familiar $p=1$ case, we write $\{f,\lambda\}:= \Pi(df,d\lambda)=\frac1{p!} \Pi^{ij_1\ldots j_p}\partial_if (d\lambda)_{j_1\ldots j_p}$
for a $(p-1)$-form~$\lambda$ and a function~$f$. In the special case, where $d\lambda$ factorizes as a product $d\lambda=d\lambda_1\wedge\cdots\wedge d\lambda_p$, we have $\{f,\lambda\} \equiv
\{f,\lambda_1,\ldots,\lambda_p\}$.
We consider a set of local coordinates $(x^{1},\dots,x^{n})$ on $M$ and denote the
corresponding indices by lower case Latin characters $i,j,k$, etc..
Upper case Latin characters $I,J,K$, etc. denote strictly
ordered $p$-tuples of indices,
i.e. $J = (j_{1}, \dots, j_{p})$ with $1\leq j_{1} < \dots <
j_{p}\leq n$. With this notation, $\Pi(df,d\lambda)=\Pi^{iJ}\partial_if (d\lambda)_J$. Often, we will omit  indices altogether, implicitly implying  matrix multiplication of the underlying rectangular matrices as in $(\Pi F^T)^i_j = \Pi^{iK} F_{Kj}$.
We use  Roman characters $a$, $B$, etc. for indices and multi-indices taking values only in the ``noncommutative'' directions $1,...,p+1$.

\section{Covariant coordinates}\label{CovX}

Before we introduce in the next section  the Nambu-Poisson gauge potential\footnote{This is the higher analog of the $p=1$ noncommutative gauge potential.} $\widehat A$ and field strength  $\widehat F$, let us define  ``covariant coordinates''\footnote{Covariant with respect to the gauge transformation (\ref{GuageTransfAI}). For $p=1$ they correspond to background independent operators of \cite{Seiberg:1999vs}; they are actually dynamical fields.} as functions $\widehat{x}^i(x)$, $i=1,\ldots, n$ of the space-time coordinates $\{x^i\}_{i=1}^{n}$, which transform
under gauge transformations parametrized by a $(p-1)$-form $ \Lambda$ as \begin{equation}\label{GuageTransfX}
\delta_{ \Lambda} \widehat x^i=\{\widehat x^i, \Lambda\} \ ,
\end{equation}
where the bracket is a $p+1$ Nambu-Poisson bracket (cf. Sec.~2 for notation).
We assume our fixed (but arbitrarily chosen) coordinates $x^i$ to be invariant under  gauge transformations. We also assume that they can be expanded around any point $x\in M$, at least in the sense of formal power series,  as $\widehat x^i=x^i + \ldots$. Hence, at least formally, we can always solve for $x^i$ as functions of covariant coordinates $\widehat x^i$, i.e. $x^i= \widehat x^i+\ldots$. We denote by $\rho$ the (formal) diffeomorphism on $M$ corresponding to this change of local variables on $M$ and write $\widehat x^i=\rho^* (x^i)$ for the respective local coordinate functions. The change of coordinates defined by $\rho^*$ is also called ``covariantizing map''. The diffeomorphism $\rho$ can be used to define a new Nambu-Poisson structure $\Pi'$ with bracket $\{\cdot,\ldots,\cdot\}'$:
\begin{equation}\label{Pi'}
\rho^*(\{ x^{i_1},\ldots, x^{i_{p+1}}\}')
:=\{\rho^* x^{i_1},\ldots, \rho^* x^{i_{p+1}}\}
\equiv \{\widehat x^{i_1},\ldots, \widehat x^{i_{p+1}}\}.
\end{equation}

\section{Nambu-Poisson gauge fields}\label{NPGF}
Here and in the subsequent sections, we follow closely the semiclassical parts of \cite{Jurco:2000fs,Jurco:2001my}, where the $p=1$ case is described. Using covariant coordinates $\widehat x^i$, we define the Nambu-Poisson (``noncommutative'') gauge potential with components labeled by upper indices $i=1,\ldots, n$ as\footnote{See \cite{Ho:2013,Ho:2013opa,Ho:2013paa} for an alternative approach related to area-preserving diffeomorphisms.}
\begin{equation}\label{Aupper}
\widehat A^i = \widehat x^i - x^i=\rho^* (x^i) - x^i.
\end{equation}
Its gauge transformation follows from (\ref{GuageTransfX})
\begin{equation}\label{GuageTransfAI}
\delta_{{\Lambda}} \widehat A^i=\{\widehat A^i, \Lambda\}+ \{x^i, \Lambda\} .
\end{equation}
Next, we introduce the contravariant tensor $F'$ with components $F'^{i_1\ldots i_{p+1}}$ as the difference of the Nambu-Poisson structures $\Pi'$, see equation (\ref{Pi'}), and $\Pi$:
\begin{equation}\label{F'upper}
F'^{i_1\ldots i_{p+1}} = \Pi'^{i_1\ldots i_{p+1}}- \Pi^{i_1\ldots i_{p+1}}.
\end{equation}
Covariantizing the individual components of this tensor using the diffeomorphism $\rho$, we obtain the Nambu-Poisson (``noncommutative'') field strength $\widehat F'$ with components
\begin{equation}\label{Fhatupper}
\widehat{F}'^{i_1\ldots i_{p+1}} := \rho^*(F'^{i_1\ldots i_{p+1}}) .
\end{equation}
Using (\ref{F'upper}) and a hat to denote the application of $\rho^*$,
\begin{equation}
\widehat{F}'^{i_1\ldots i_{p+1}} = \widehat \Pi'^{i_1\ldots i_{p+1}} - \widehat{\Pi}^{i_1\ldots i_{p+1}} = \rho^*(\Pi'^{i_1\ldots i_{p+1}}) - \rho^*(\Pi^{i_1\ldots i_{p+1}}) .
\end{equation}
Rewriting this with the help of (\ref{Pi'}) as
\begin{equation}
\widehat F'^{i_1\ldots i_{p+1}}=\{\widehat x^{i_1},\ldots, \widehat x^{i_{p+1}}\}-\{x^{i_1},\ldots, x^{i_{p+1}}\}(\widehat{x}) ,
\end{equation}
the gauge transformation of $\widehat F'$ can be easily  determined:
\begin{equation}\label{GuageTransfFI}
\delta_{{\Lambda}} \widehat F'^{i_1\ldots i_{p+1}}=\{\widehat F'^i, \Lambda\}  .
\end{equation}

From now on we will assume without loss of generality that the local coordinates $x^i$ are adapted to the Nambu-Poisson structure $\Pi$, i.e., $\{x^i\}$ are local coordinates around some point $M$, where $\Pi$ is non-zero, such that\footnote{Here we ignore, for simplicity,  points where $\Pi$ could possibly be zero and focus on globally non-degenerate Nambu-Poisson structures.}
\begin{equation} \label{Pi_partials}
\Pi= \partial_1\wedge\dots\wedge \partial_{p+1}.
\end{equation}
With this choice of coordinates, we find
\begin{equation}\label{Fupper_constantNambu}
\widehat F'^{i_1\ldots i_{p+1}}=\{\widehat x^{i_1},\ldots, \widehat x^{i_{p+1}}\}-\{x^{i_1},\ldots, x^{i_{p+1}}\},
\end{equation}
where the second bracket is in fact either zero or equal to the $p+1$ epsilon symbol in the noncommutative directions $1,\ldots,p+1$. Roman indices $a_1,\ldots, a_{p+1}$ shall henceforth  denote these directions.
Furthermore, we will focus on the case where for the covariantizing map $\rho^*$ acts trivially (i.e. $\widehat x^i =x^i$)  on coordinates
$x^i$ with indices in the commutative directions $p+2,\ldots , n$.
It follows from (\ref{GuageTransfX}) that only the covariant coordinates in the noncommutative directions transform non-trivially under gauge transformations and that the gauge fields $\widehat{A}^i$ are trivial for $i=p+2,\ldots ,n$. Also, all the field strengths, except those indexed solely by noncommutative indices $i=1,\ldots,p+1$, will automatically be zero.
With these conventions, we can use the $p+1$ epsilon tensor to lower the index on $\widehat A^a$ and introduce another kind of gauge potential uniquely determined by complete antisymmetrization of its non-zero components $\widehat A_B$ labeled by strictly ordered $p$-tuples of indices, with individual indices taking values in the labels of the noncommutative directions \begin{equation}\label{Alower}
\widehat A_{B}:= \epsilon_{aB}\widehat A^a.
\end{equation}
The components $\widehat A_B$ transform in a more familiar looking manner (but recall that we are still dealing with a $p+1$ Nambu-Poisson  bracket and a $(p-1)$-form gauge parameter $\Lambda$):
\begin{equation}\label{GuageTransfAIII}
\delta_{ \Lambda} \widehat A_{B} =(d\Lambda)_B + \{\widehat A_{B},\Lambda\} .
\end{equation}
Similarly, we define the corresponding field strength with components $\widehat F_{aB}'$ by
\begin{equation}\label{FlowerBullet}
\widehat F'_{aB} = \epsilon_{aC}(\widehat \Pi'^{bC} - {\Pi}^{bC})\epsilon_{bB}.
\end{equation}
The components $\widehat  F'_{aB}$ transform as expected
\begin{equation}\label{GuageTransfFII}
\delta_{ \Lambda} \widehat  F'_{aB} =\{\widehat F'_{aB},\Lambda\}.
\end{equation}
A straightforward check reveals that $\widehat F'_{aB}$ can be consistently extended to be antisymmetric in all of its indices.
Finally, $\widehat  F'_{aB}$  can be expressed in terms of the gauge potential $\widehat{A}_{B}$. For this, we need to  a $(p+1-q)$-ary Nambu bracket defined as\footnote{With some abuse of notation we allow also for the case $p=q$, i.e., the ``1-ary" bracket, which will become useful later.}
\[ \{ \cdot, \dots, \cdot \}^{i_{1} \dots i_{q}} := \{ x^{i_{1}}, \dots, x^{i_{q}}, \cdot, \dots, \cdot \}. \]
Now, using (\ref{Aupper}), (\ref{Fupper_constantNambu}), (\ref{Alower}) and (\ref{FlowerBullet}) we obtain
\begin{equation} \label{Fprime_As}
\widehat{F}'_{1 \dots p+1} = (d\widehat{A})_{ 1 \dots p+1} + \sum_{r=0}^{p-1} \sum_{\sigma\in S(r,n-r)} (-1)^{\sum_{k=r+1}^{p+1} (\sigma(k)-1)} {\mbox{sgn}}(\sigma)\{ \widehat{A}_{[\sigma(r+1)]}, \dots ,\widehat{A}_{[\sigma(p+1)]} \}^{\sigma(1)\dots\sigma(r)},
\end{equation}
where $\sigma\in S(r,n-r)$ is an $(r,n-r)$ shuffle, and $[a]$ is the multi-index $1\cdots (a-1)(a+1)\cdots (p+1)$. This formula is a generalization to $p>1$ of the well-known $p=1$ formula for the (noncommutative) field strength that involves the 2-bracket (``commutator'') of gauge fields.

In the next section we will use a higher analog of the Seiberg-Witten map in order to construct explicit expressions for the covariant coordinates and noncommutative gauge fields. This will allow us to also supplement the remaining components of the Nambu-Poisson gauge field strength (\ref{FlowerBullet}), i.e., the ones with at least one index in a commutative direction.

\section{Nambu-Poisson gauge fields via Seiberg-Witten map}\label{via SW}
We start with a brief summary of the relevant facts concerning the Seiberg-Witten map as it applies in the present context. We refer the reader to a detailed exposition in \cite{JSV1}. All order solution to the Seiberg-Witten map related to Nambu-Poisson M5-brane theory can be found in \cite{Chen:2010br}. 

Let us consider a $p$-form gauge potential $a$ on $M$ with corresponding field strength $F=da$. Infinitesimally, under a gauge transformation given by a $(p-1)$-form $\lambda$,
\begin{equation}
\delta_\lambda a = d\lambda, \hskip 0.5cm \delta_\lambda F = 0.
\end{equation}
Using the $(p+1)$-form $F$ we construct from a given Nambu-Poisson tensor $\Pi$ the
$F$-gauged tensor which we denote for now by $\Pi_F$,\footnote{We assume that $1 - \Pi F^{T}$ is invertible. In a more formal approach we also could treat $\Pi_F$ as a formal power series in $\Pi$.\label{invertible}.}
\begin{equation}
\Pi_F := (1 - \Pi F^{T})^{-1} \Pi=\Pi(1 - F^{T}\Pi)^{-1}.
\end{equation}
These expressions are to be interpreted as matrix equations for the corresponding maps sending $p$-forms to 1-forms, cf. Sec. 2. The superscript $T$ stands for the transposed map. For $p>1$, the $(p+1)$-tensor $\Pi_F$ is always a Nambu-Poisson one,\footnote{Even for a non-closed $F$.}  furthermore, we also have due to factorizability of $\Pi$,
\begin{equation}  \label{eq_thetaprimetrace}
\Pi_F = \left(1 - \frac{1}{p+1}\< \Pi, F\>\right)^{-1} \Pi,
\end{equation}
where $\<\Pi,F\> = \Pi^{iJ} F_{iJ} \equiv \Tr(\Pi F^{T})$.

Now we define a $1$-parametric family of Nambu-Poisson tensors
$\Pi_{t} := (1 - t\Pi F^{T})^{-1} \Pi$, cf. Footnote \ref{invertible}, interpolating between $\Pi$ and $\Pi_F$.
Differentiation of $\Pi_{t}$ with respect to $t$ gives:
\begin{equation}
\partial_{t} \Pi_{t} = \Pi_{t} F^{T} \Pi_{t}.
\end{equation}
This equation can be rewritten as
\begin{equation} \label{eq_difffortheta}
\partial_{t} \Pi_{t} = -\mathcal{L}_{A^{\sharp}_{t}} \Pi_{t},
\end{equation}
where the time-dependent vector field $A^{\sharp}_{t}$ is defined as
$A^{\sharp}_{t} = {\Pi}^{\sharp}_{t}(a)  = {\Pi}^{iJ}_{t} a_J \partial_i$ and $\mathcal{L}_{A^{\sharp}_{t}}$ is the corresponding Lie derivative. Equation
(\ref{eq_difffortheta}) implies that the flow $\phi_{t}$
corresponding to $A_{t}^{\sharp}$, together with the initial condition
$\Pi_{0} = \Pi$, maps $\Pi_{t}$ to $\Pi$, that is,
\begin{equation}
\phi_{t}^{\ast}(\Pi_{t}) = \Pi.
\end{equation}
We have thus found the map $\rho_{a} := \phi_{1}$, such that
$\rho_{a}^{\ast}(\Pi') = \Pi$. This is the higher form gauge field ($p > 1$) analogue
of the well known semiclassical Seiberg-Witten map. We emphasize the
dependence of this map on the $p$-form $a$ by an explicit addition
of the subscript $a$.
The following observation is important: The Nambu-Poisson tensor $\Pi_t$ is gauge invariant (because it depends on the $p$-potential $a$ only via the gauge invariant $p+1$ form field strength $f=da$), but the Nambu-Poisson map $\rho_{a}$ is not: The infinitesimal gauge transformation $\delta_\lambda a = d \lambda$, with a
$(p-1)$-form gauge transformation parameter $\lambda$, induces a change in the flow, which is generated by the vector field $X_{[\lambda, a]} = \Pi^{iJ} d\Lambda_J\partial_i$, where the $(p-1)$-form $\Lambda$, explicitly given by
\begin{equation}
\Lambda= \sum_{k=0}^{\infty} \frac{(\mathcal{L}_ {A_{t}^\sharp} +
\partial_t)^k(\lambda)}{(k+1)!}\Big| _{t=0}  \,,
\end{equation}
is the semiclassically noncommutative $(p-1)$-form gauge parameter. This leads to the following rule for the gauge transformation of coordinates $\widehat{x}_a^i:=\rho_a^\ast (x^i)$, cf. (\ref{GuageTransfX}):
\begin{equation}
\delta_\lambda\widehat{x}_a^i=\{\widehat x_a^i,\Lambda\}.
\end{equation}
Hence, the generalized Seiberg-Witten map provides us with an explicit construction, based on ordinary higher gauge fields, of the covariant coordinates $\widehat x^i$ that we introduced in Sec. \ref{CovX}. As a consequence, we can identify $\widehat x^i\equiv \widehat{x}^i_a$ and $\Pi'\equiv \Pi_F$. Moreover, $\widehat x^i = \widehat{x}_a^i = x^i$, for the ``commutative'' directions $i=p+2,\ldots n$. All discussion of the previous sections \ref{CovX} and \ref{NPGF} applies directly.

Having the ordinary $p$-form gauge field $a$ at our disposal we can now define
the full Nambu-Poisson field strength $\widehat{F}'$ with all components (in noncommutative as well as in commutative directions), such that that its components in the noncommutative directions $x^1,\dots , x^{p+1}$  coincide with those of $\widehat{F}_{aB}'$ (\ref{FlowerBullet}).

For this let
\begin{eqnarray} \label{def_Fprime}
F':= F(1-\Pi^TF)^{-1}= (1-F\Pi^T)^{-1}F
\end{eqnarray}
and define
\begin{equation}\label{NCF}
\widehat F'_{iJ} :=\rho^{\ast}_A F'_{iJ},
\end{equation}
i.e., the components of $F'$ evaluated in the covariant coordinates.
It is a rather straightforward check to see that for all indices $i_1,\ldots ,i_{p+1}$ taking  values only in the set $\{1,\ldots ,p+1\}$ we get exactly the $\widehat{F}'_{aB}$ of (\ref{FlowerBullet}).

Now we turn our attention to the remaining components of $\widehat{F}'$ (including commutative directions).
Starting from (\ref{def_Fprime}) and (\ref{NCF}), we can with the help of (\ref{invertible}) and the explicit expression for $\Pi$ in coordinates (\ref{Pi_partials}) use a construction very similar to the one leading to (\ref{Fprime_As}). We find that the resulting expressions involve a covariant scalar function that depends on $\widehat A$ (and hence via the generalized Seiberg-Witten map also on the ordinary p-form gauge potential $a$):
\[
f[\widehat A]:=1+ \sum_{r=0}^{p} \sum_{\sigma\in S(r,n-r)} (-1)^{\sum_{k=r+1}^{p+1} (\sigma(k)-1)} {\mbox{sgn}}(\sigma)\{ \widehat{A}_{[\sigma(r+1)]}, \dots, \widehat{A}_{[\sigma(p+1)]} \}^{\sigma(1)\dots\sigma(r)}.
\]
Firstly, let us consider $\widehat{F}_{aK}'$ with the index $a$ taking on  values in $\{1,\dots,p+1\}$, and $K$ containing at least one index in one of the commutative directions $p+2,\dots,n$.
We find
\begin{equation}
\widehat{F}_{aK}'= f[\widehat A]\widehat{F}_{aK},
\end{equation}
where $\widehat{F}_{aK} = \rho^*F_{aK}$ is the component $F_{aK}$ of the ordinary (commutative) field strength evaluated at the covariant coordinates $\widehat x^i$.
Secondly, for the components of $\widehat F'$ with index $k$ taking value in $\{p+2,\dots,n\}$, and $A$ containing only the indices lying in the set $\{1, \dots, p+1\}$,
\begin{equation}
\widehat{F}_{kA}'= f[\widehat A]\widehat{F}_{kA},
\end{equation}
Finally, for the components $\widehat{F}_{kL}'$, where $k$ takes value in the set $\{p+2, \dots, n\}$ and $L$ contains at least one index of the same set, we have
\begin{equation}
\widehat{F}'_{kL} = \widehat{F}_{kL}+f[\widehat A] \sum_{a=1}^{p+1} (-1)^{a+1} \widehat{F}_{k[a]} \widehat{F}_{aL}.
\end{equation}
Under (ordinary) infinitesimal gauge transformations $\delta_\lambda$, all components of $\widehat F'$ transform as
\begin{equation}\label{NCF2}
\delta_\lambda\widehat F' =\{\widehat F',\Lambda\},
\end{equation}
justifying calling it ``Nambu-Poisson'' or ``(semiclassically) noncommutative" field strength.

Note that unlike for the noncommutative components, the full tensor $\widehat F'$ cannot be extended to be a totally antisymmetric one.

\section{Action}\label{action}

For simplicity, we assume  Euclidean space-time signature.\footnote{Another simple possibility would be consider the Minkowskian space-time, with $\Pi$ extending in the spatial directions only. In case of a general metric $g$ we would have to use the inverse metric matrix elements evaluated in the covariant coordinates to rise the indices of $\hat F'$ and the density defined by the metric also evaluates in the covariant coordinates.}
The action
\begin{equation} \label{def_action}
\frac1{g}\int_M d^nx \widehat F'_{iJ}\widehat F'^{iJ}
\end{equation}
is by construction invariant under ordinary commutative as well as under Nambu-Poisson (semiclassically noncommutative) gauge transformations. This can easily be verified using partial integration. The coupling constant $g$ is dimensionless in $n = 2(p+1)$ spacetime dimensions, i.e. for example for $p=1$, $n=4$ (NC Maxwell) and for $p=2$, $n=6$ (M2-M5 system). In the following we will set $g=1$.

We  expand $\widehat{F}'$ in a power series in $\Pi$
\begin{equation}
\widehat{F}'_{iJ} = F_{iJ} + A_{L} \Pi^{kL} F_{iJ,k} + F_{iL} \Pi^{kL} F_{kJ} + o(\Pi^{2}).
\end{equation}
The corresponding expansion of the action (\ref{def_action}) is
\begin{equation}\label{expansion}
\int_M d^nx \widehat F'_{iJ}\widehat F'^{iJ} = \int_{M} d^{n}x \left\{ F_{iJ} F^{iJ} - \frac{1}{p+1} F_{iJ}F^{iJ} F_{kL} \Pi^{kL} + 2 F^{iJ} F_{iL} \Pi^{kL} F_{kJ} \right\} + o(\Pi^{2}).
\end{equation}
A quantization of the underlying Nambu-Poisson structure will not add quantum corrections to the action at the given order of expansionr.

Shifting the components $\widehat{F}'_{1\ldots p+1}$ of the Nambu-Poisson field strength by the constants $\epsilon_{1\ldots p+1}$, will not affect the gauge invariance of the the action (\ref{def_action}). Using (\ref{Fupper_constantNambu}) and (\ref{FlowerBullet}) we see that the  action (\ref{def_action}) with shifted $\widehat F'$ takes the form of a semiclassical version of a Nambu-Poisson matrix model:
\begin{equation}
S_M=\int d^nx  \{ \widehat{x}^{a}, \widehat{x}^{A}\} \{ \widehat{x}_{a},\widehat{x}_{A} \} = \int d^nx \frac{1}{p!} \{ \widehat{x}^{a_{1}}, \dots, \widehat{x}^{a_{p+1}} \} \{ \widehat{x}_{a_{1}}, \dots, \widehat{x}_{a_{p+1}} \},
\end{equation}
where the summation in the second expression runs over all (not strictly ordered) $(p+1)$-indices $(a_{1}, \dots, a_{p+1})$ and $(b_{1}, \dots, b_{p+1})$, with all of them in the noncommutative direction. We could actually drop the a priori restriction of the summation to noncommutative directions, since the Nambu-Poisson bracket automatically takes care of this. For a more detailed discussion of the (semiclassical) matrix model we refer to \cite{JSV1}.

Given an appropriate quantization $[\cdot,\ldots,\cdot]$ of the Nambu-Poisson bracket and trace of the quantized Nambu-Poisson structure, the Nambu-Poisson matrix model takes the form
\begin{equation}
\widetilde S_M = \frac{1}{p!}\Tr  [ \widehat{x}^{a_{1}}, \dots, \widehat{x}^{a_{p+1}} ][ \widehat{x}_{a_{1}}, \dots, \widehat{x}_{a_{p+1}} ] .
\end{equation}
There have been several attempts to find a consistent quantization of Nambu-Poisson structures. One of these \cite{Mylonas:2012pg} is in fact suitable for our purposes (at least in the case $p=2$): It is an approach based on nonassociative star product algebras on phase space, whose Jacobiator defines a quantized Nambu-Poisson bracket on configuration space. Let us mention without going into details that this approach can be adapted to provide a consistent quantization of the Nambu-Poisson gauge theory described in this letter, including a quantization of the generalized Seiberg-Witten maps. Details of this construction are beyond the scope of the present letter and will be reported elsewhere.

\section*{Acknowledgements}
It is a pleasure to thank Tsuguhiko Asakawa, Peter Bouwknegt, Chong-Sun Chu, Pei-Ming Ho, Petr Ho\v rava, Dalibor Kar\'asek, Noriaki Ikeda, Matsuo Sato, and Satoshi Watamura for helpful discussions. B.J. and P.S. appreciate the hospitality of the Center for Theoretical Sciences, Taipei, Taiwan, R.O.C.. B.J. thanks CERN for hospitality.
We gratefully acknowledge support by grant GA\v CR P201/12/G028 (B.J.), by the DFG within the Research Training Group 1620 ``Models of Gravity'' (J.V., P.S.), and by the Grant Agency of the Czech Technical University in Prague, grant No. SGS13/217/OHK4/3T/14 (J.V.). We thank the DAAD (PPP) and ASCR \& MEYS (Mobility) for supporting our collaboration.

\section*{References}


\end{document}